\documentstyle[12pt,epsfig,graphicx,amsmath,amsthm]{article} 
\makeatletter \def\@cite#1#2{{[{#1}]\if@tempswa\typeout {IJCGA
warning: optional citation argument ignored: `#2'} \fi}}

\newcount\@tempcntc
\def\@citex[#1]#2{\if@filesw\immediate\write\@auxout{\string\citation{#2}}\fi
  \@tempcnta\z@\@tempcntb\m@ne\def\@citea{}\@cite{\@for\@citeb:=#2\do
    {\@ifundefined
       {b@\@citeb}{\@citeo\@tempcntb\m@ne\@citea\def\@citea{,}{\bf ?}\@warning
       {Citation `\@citeb' on page \thepage \space undefined}}%
    {\setbox\z@\hbox{\global\@tempcntc0\csname b@\@citeb\endcsname\relax}%
     \ifnum\@tempcntc=\z@ \@citeo\@tempcntb\m@ne
       \@citea\def\@citea{,}\hbox{\csname b@\@citeb\endcsname}%
     \else
      \advance\@tempcntb\@ne
      \ifnum\@tempcntb=\@tempcntc
      \else\advance\@tempcntb\m@ne\@citeo
      \@tempcnta\@tempcntc\@tempcntb\@tempcntc\fi\fi}}\@citeo}{#1}}
\def\@citeo{\ifnum\@tempcnta>\@tempcntb\else\@citea\def\@citea{,}%
  \ifnum\@tempcnta=\@tempcntb\the\@tempcnta\else
   {\advance\@tempcnta\@ne\ifnum\@tempcnta=\@tempcntb \else 
\def\@citea{--}\fi
    \advance\@tempcnta\m@ne\the\@tempcnta\@citea\the\@tempcntb}\fi\fi}
\makeatother

\def\boxit#1{\leavevmode\thinspace\hbox{\vrule\vtop{\vbox{\hrule%
        \vskip3pt\kern1pt\hbox{\vphantom{\bf/}\thinspace\thinspace%
        {\bf#1}\thinspace\thinspace}}\kern1pt\vskip3pt\hrule}\vrule}%
        \thinspace}
\def\Boxit#1{\noindent\vbox{\hrule\hbox{\vrule\kern3pt\vbox{
\advance\hsize-7pt\vskip-\parskip\kern3pt\bf#1 \hbox{\vrule height0pt
depth\dp\strutbox width0pt} \kern3pt}\kern3pt\vrule}\hrule}}

\newcommand{\gsim}{\lower.7ex\hbox{$\;\stackrel{\textstyle>}{\sim}\;$}}
\newcommand{\lsim}{\lower.7ex\hbox{$\;\stackrel{\textstyle<}{\sim}\;$}}

\allowdisplaybreaks[1]

\def\bad{\begin{aligned}[t]}
\def\ead{\end{aligned}}

\def\ifmath#1{\relax\ifmmode #1\else $#1$\fi}

\def\sM{\ifmath{{\bf M}}}

\def\mLs{\ifmath{{{\bf m}_L^2}}}
\def\mes{\ifmath{{{\bf m}_e^2}}}
\def\mns{\ifmath{{{\bf m}_\nu^2}}}

\def\mHus{\ifmath{m_{H_u}^2}}

\def\Ann{\ifmath{A_\nu}}
\def\Aen{\ifmath{A_e}}

\def\Yn{\ifmath{{{\bf Y}_\nu}}}
\def\Ye{\ifmath{{{\bf Y}_e}}}

\def\An{\ifmath{{{\bf A}_\nu}}}
\def\Ae{\ifmath{{{\bf A}_e}}}

\def\tra{\ifmath{{\!\!\!T}}}

\def\mass{\ifmath{{\cal M}}}
\def\hatmass{\ifmath{\widehat{\cal M}}}
\def\mtilde1{\ifmath{\widetilde m_1}}
\def\mtildea{\ifmath{\widetilde m_{1\alpha}}}

\begin{document}
%
%

\begin{titlepage}

\title{\bf  Probing Supersymmetric Leptogenesis with $\mu\rightarrow e\gamma$}
\vskip3in \author{{\bf Alejandro Ibarra} and
{\bf Cristoforo Simonetto\footnote{\baselineskip=16pt {\small E-mail addresses: {\tt
alejandro.ibarra@ph.tum.de, cristoforo.simonetto@ph.tum.de}}}}
\hspace{3cm}\\ \vspace{0.1cm}
{\normalsize\it  Physik-Department T30d, Technische Universit\"at M\"unchen,}\\[-0.05cm]
{\normalsize \it James-Franck-Strasse, 85748 Garching, Germany.}
}  \date{}  \maketitle  \def\baselinestretch{1.15}
\begin{abstract}
\noindent

Extending the Minimal Supersymmetric Standard Model with three
right-handed neutrino superfields is one of
the best motivated scenarios for physics beyond the Standard Model.
However, very little is known from observations about the high
energy parameters of this model.
In this paper we show, under the plausible assumptions that the
neutrino Yukawa eigenvalues are hierarchical and the absence
of cancellations, that there exists 
an upper bound on the smallest Yukawa eigenvalue stemming 
from the non-observation of the rare lepton decay $\mu\rightarrow e \gamma$.
Furthermore, we show that this bound implies an upper bound
on the lightest right-handed neutrino mass of approximately
$5\times 10^{12}$ GeV for  typical supersymmetric parameters.
We also discuss the implications of this upper bound
for the minimal leptogenesis scenario based on the decay
of the lightest right-handed neutrino and
we argue that an improvement of sensitivity of six orders
of magnitude to the process $\mu\rightarrow e\gamma$ 
could rule out this mechanism as the origin
of the observed baryon asymmetry, unless the neutrino
parameters take very specific values.

\end{abstract}

\thispagestyle{empty}
\vspace*{0.2cm} \leftline{June 2009} \leftline{}

\vskip-18.0cm \rightline{TUM-HEP 713/09}

\vskip3in

\end{titlepage}
\setcounter{footnote}{0} \setcounter{page}{1}
\newpage
\baselineskip=20pt

\noindent

\section{Introduction}

One of the simplest and best motivated extensions of the Standard
Model consists on adding to the particle content three right-handed neutrinos. 
Being singlets under the Standard Model gauge group, the most
general Lagrangian has to include not only a Yukawa coupling with the
lepton and the Higgs doublets, but also a Majorana mass term 
for the right-handed neutrinos. If, after the electroweak symmetry
breaking, the right-handed Majorana masses are much larger than 
the Dirac neutrino masses, the active neutrinos will acquire
effective masses
which are much smaller than the electroweak symmetry breaking scale.
This is the renowned see-saw mechanism~\cite{seesaw}.

Although very appealing theoretically, the see-saw mechanism
faces the disadvantage of lacking predictability. The see-saw
mechanism makes the definite prediction of the existence
of neutrino masses and strongly suggests the existence of CP violation in 
the lepton sector. However, it cannot predict the concrete values of the
neutrino masses, the mixing angles or the CP violating phases. 
Besides predicting non-vanishing neutrino masses,
the see-saw mechanism makes a second definite prediction.
If the interactions of the right-handed neutrinos with the lepton and Higgs
doublets do not preserve CP, the out of equilibrium decays of the right-handed 
neutrinos in the primeval plasma will generate a baryon asymmetry through
the leptogenesis mechanism~\cite{Fukugita:1986hr}, 
provided the mass of the lightest right-handed neutrino 
is larger than $\sim 100$ GeV, which is the temperature
below which the sphaleron interactions can no longer convert 
the generated lepton asymmetry into a baryon asymmetry.
Remarkably, the out of equilibrium decays of the right-handed
neutrinos could account for {\it all} the observed baryon asymmetry 
if the mass of the lightest right-handed neutrino is larger than 
$10^9$ GeV~\cite{Davidson:2002qv,Buchmuller:2004nz}.
However, the large right-handed neutrino masses required by the
leptogenesis mechanism preclude any hope to test directly the 
see-saw mechanism. 

On the other hand, the existence of such heavy particles interacting
with the Higgs doublet strongly suggests the existence of 
supersymmetry (SUSY), in order to protect the Higgs mass against
quadratically divergent quantum corrections. In the supersymmetric
version of the see-saw mechanism, the flavour and CP violating 
effects of the neutrino Yukawa coupling typically propagate to
the soft SUSY breaking terms through quantum 
corrections~\cite{Borzumati:1986qx}, thus reopening the possibility
of probing this interesting scenario through lepton flavour
violating processes, such as $\mu\rightarrow e \gamma$, 
or through leptonic electric dipole moments. 

Unfortunately, this new opportunity to probe the see-saw mechanism
is hindered by the large number of parameters in the high energy
Lagrangian. The complete leptonic Lagrangian depends on fifteen real
parameters and six phases, of which only nine real parameters and
three phases are in principle accessible at low energies (three
charged lepton masses, three neutrino masses, three mixing angles
and three CP phases). Furthermore, it can be shown that the see-saw
mechanism can accommodate any observed rates for
the rare lepton decays or the electric dipole moments, while being
consistent with the observed neutrino parameters. Namely, there 
exists a one to one correspondence between the high energy
see-saw parameters and the combinations of Yukawa couplings and
right-handed masses which are relevant to low energy 
experiments~\cite{Davidson:2001zk}.
In consequence, it is {\it impossible} to make any completely model 
independent prediction about the see-saw mechanism. Nevertheless, 
any assumption about the high-energy see-saw parameters will
break this one-to-one correspondence and will lead to constraints
among the low energy parameters, or well defined relations between
the high energy see-saw parameters and observable quantities. 
Several works have appeared in the literature aiming to derive
from laboratory experiments constraints on the see-saw parameters,
either working in specific high-energy frameworks or pursuing a more
phenomenological approach, where the constraints somehow rely on
additional assumptions on the high-energy 
Lagrangian~\cite{general,Masiero:2002jn}. 
Other works have aimed to find connections between leptogenesis and
observable quantities, such as the rates for the rare lepton decays
or the leptonic electric dipole 
moments,~\cite{Davidson:2002em,Davidson:2003cq,low-lepto},
again imposing conditions on the high-energy theory. 

The most minimal assumption that one could make on the high-energy
see-saw parameters is the absence of artificial cancellations among
terms when computing the low energy predictions. It is remarkable
that this minimal assumption already leads to a correlation among
low energy observables of the form ${\rm BR}(\mu\rightarrow e\gamma)\gsim
C \times {\rm BR}(\tau\rightarrow \mu\gamma){\rm BR}(\tau\rightarrow e\gamma)$,
where $C$ is a constant that depends on supersymmetric 
parameters~\cite{Ibarra:2008uv,Ibarra:2008uq}.
It was shown in~\cite{Ibarra:2008uv} that if present $B$-factories
 discover both
$\tau\rightarrow \mu\gamma$ and $\tau\rightarrow e\gamma$, the see-saw
mechanism would be ruled out in large regions of the SUSY parameter
space (assuming universal boundary conditions at the Grand Unification
scale). This result is a proof of principle that popular supersymmetric
scenarios incorporating the see-saw mechanism could be ruled out
using low energy experiments, with the only assumption of the lack
of cancellations among parameters.

In this paper we would like to explore the implications of
another well motivated assumption about the high energy
see-saw parameters, namely that the neutrino Yukawa eigenvalues
are hierarchical. Our motivation to consider
this scenario is the observation of large hierarchies in
the eigenvalues of {\it all} known Yukawa matrices.
Therefore, although the particular mechanism which generates the Yukawa
couplings is completely unknown, observations suggest that this putative
mechanism prefers to generate Yukawa matrices with hierarchical eigenvalues.

In Section 2 we will review a parametrization of the neutrino Yukawa
couplings which will prove to be useful in studying the
implications of the see-saw mechanism for the rare decays,
under the assumption of hierarchical neutrino Yukawa
eigenvalues. In Section 3 we will derive lower bounds on the rates of
the leptonic rare decays as a function of the eigenvalues of the Yukawa
couplings and neutrino parameters, and we will use the experimental 
bounds on the rare decays to derive constraints on the neutrino Yukawa
eigenvalues. In Section 4 we will discuss the implications of
these bounds on the leptogenesis mechanism to generate the baryon 
asymmetry of the Universe. Lastly, in Section 5 we will present our conclusions.
\section{The see-saw mechanism with hierarchical Yukawa eigenvalues}

In the supersymmetric see-saw mechanism, the particle content of
the Minimal Supersymmetric Standard Model (MSSM) is extended
with three right-handed neutrino superfields, ${\nu_{R}}_i$, $i=1,2,3$,
singlets under the Standard Model gauge group.
Imposing $R$-parity conservation, the leptonic superpotential reads:
\begin{equation}
W_{\rm lep}= {e_R^c}_i \Ye_{ij} L_j  H_d+
{\nu_R^c}_i \Yn_{ij} L_j  H_u
- \frac{1}{2}{\nu_R^c}_i{\sM}_{ij}{\nu_R^c}_j \;,
\label{superp} 
\end{equation}
where $H_u$ and $H_d$ are the hypercharge $+1/2$ and $-1/2$ Higgs doublets,
respectively, $\Ye$ and $\Yn$ are the matrices of charged lepton and 
neutrino Yukawa couplings, respectively, 
and  $\sM$ is a $3\times 3$ Majorana mass matrix. 
It is natural to assume that the right-handed neutrino masses
are much larger than the 
electroweak scale or any soft mass. If this is the case, 
the theory is well described at low energies by the following 
effective superpotential:
\begin{equation} 
W^{\rm eff}_{\rm lep}={e_R^c}_i \Ye_{ij} L_j  H_d+\frac{1}{2}
\left(\Yn^\tra{\sM}^{-1}\Yn \right)_{ij} (L_i H_u)(L_j H_u)\;,
\label{effsuperp}
\end{equation}
which generates neutrino masses after the electroweak symmetry
breaking.
In the phenomenological studies it is convenient to work in the
leptonic basis where the charged lepton
Yukawa coupling and the right-handed Majorana mass matrix are real and
diagonal, namely $\Ye={\rm diag}(y_e, y_\mu, y_\tau)$ and 
$\sM={\rm diag}(M_1,M_2,M_3)\equiv D_M$,
with $M_1\leq M_2 \leq M_3$.
Then, in this basis, the neutrino mass matrix is given by
\begin{equation}
{\cal M}=\Yn^\tra D_M^{-1}\Yn\; \langle H_u^0\rangle^2 \;,
\label{mass-matrix}
\end{equation}
where $\langle H_u^0\rangle=v\, \sin\beta$ and $v=174$ GeV.
The neutrino mass matrix can be diagonalized by a unitary matrix $U$ 
yielding $U^T {\cal M} U ={\rm diag}(m_1,m_2,m_3)$, being the eigenvalues,
$m_i$, naturally very small due to
the suppression by the large right-handed neutrino mass scale.

We will work throughout this paper under the assumption that
the neutrino Yukawa coupling has  hierarchical eigenvalues.
Therefore, it is convenient to parametrize the neutrino Yukawa
coupling using the familiar singular value decomposition 
$\Yn = V_R D_Y V_L^\dagger$, where $V_R$ and $V_L$ are $3\times 3$
unitary matrices and $D_Y\equiv{\rm diag}(y_1, y_2, y_3)$ is
the diagonal matrix of eigenvalues of the Yukawa coupling
(with the convention $y_1\leq y_2 \leq y_3$). 

Substituting this parametrization in Eq.~(\ref{mass-matrix}) we find
\begin{equation}
{\cal M}=V_L^* D_Y V_R^T D^{-1}_M V_R D_Y V_L^\dagger\langle H_u^0\rangle^2\;,
\end{equation}
from where the right-handed neutrino parameters $D_M$ and $V_R$
can be calculated in terms of the measurable 
neutrino mass matrix and the parameters $D_Y$ and $V_L$. To
this end, we first rewrite the previous equation as
\begin{equation}
V_R^T D^{-1}_M V_R=
\frac{1}{\langle H_u^0\rangle^2} D^{-1}_Y V_L^T \mass V_L D^{-1}_Y=
\frac{1}{\langle H_u^0\rangle^2}D^{-1}_Y \hatmass D^{-1}_Y\;,
\label{VMV}
\end{equation}
where we have defined for convenience $\hatmass\equiv V_L^T \mass V_L$.
Then, barring cancellations and assuming a large hierarchy 
among the neutrino Yukawa eigenvalues, it follows that
\begin{equation}
V_R^\dagger D^{-2}_M V_R =\frac{1}{\langle H_u^0\rangle^4}
D^{-1}_Y \hatmass^\dagger  D^{-2 }_Y \hatmass D^{-1}_Y\simeq 
\frac{1}{\langle H_u^0\rangle^4}
\frac{1}{y^2_1} 
\begin{pmatrix}
\frac{|\hatmass_{11}|^2}{y^2_1}&
\frac{\hatmass^*_{11}\hatmass_{12}}{y_1 y_2} &
\frac{\hatmass^*_{11}\hatmass_{13}}{y_1 y_3} \\
\frac{\hatmass^*_{12}\hatmass_{11}}{y_1 y_2} &
\frac{|\hatmass_{12}|^2}{y^2_2} &
\frac{\hatmass^*_{12}\hatmass_{13}}{y_2 y_3}\\
\frac{\hatmass^*_{13}\hatmass_{11}}{y_1 y_3} &
\frac{\hatmass^*_{13}\hatmass_{12}}{y_2 y_3} &
\frac{|\hatmass_{13}|^2}{y^2_3} 
\end{pmatrix}\;,
\end{equation}
from where it is straightforward to extract the smallest
right-handed neutrino mass, $M_1$. On the other hand, taking the
inverse of Eq.~(\ref{VMV}), the same set of assumptions
leads to:
\begin{eqnarray}
V_R^\dagger D^{2}_M V_R &=& \langle H_u^0\rangle^4
D_Y \hatmass^{-1}  D^{2}_Y (\hatmass^{-1})^{\dagger} D_Y  \nonumber \\
&\simeq& \langle H_u^0\rangle^4 y_3^2
\begin{pmatrix}
y^2_1|\hatmass^{-1}_{13}|^2&
y_1 y_2\hatmass^{-1*}_{23}\hatmass^{-1}_{13} &
y_1 y_3\hatmass^{-1*}_{33}\hatmass^{-1}_{13} \\
y_1 y_2\hatmass^{-1*}_{13}\hatmass^{-1}_{23}&
y^2_2|\hatmass^{-1}_{23}|^2 &
y_2 y_3\hatmass^{-1*}_{33}\hatmass^{-1}_{23}\\
y_1 y_3\hatmass^{-1*}_{13}\hatmass^{-1}_{33} &
y_2 y_3 \hatmass^{-1*}_{23}\hatmass^{-1}_{33} &
y^2_3|\hatmass^{-1}_{33}|^2
\end{pmatrix}\;,
\end{eqnarray}
from where the largest right-handed neutrino mass, $M_3$, can be extracted.
Lastly, from taking the determinant of  Eq.~(\ref{VMV}) the intermediate
eigenvalue, $M_2$, can be derived. The approximate expressions for
the three right-handed neutrino masses 
are~\cite{Davidson:2002em,Davidson:2003cq,Branco:2002kt,Akhmedov:2003dg}:
\begin{align}
\label{RH-masses}
M_1 &\simeq y_1^2 \langle H_u^0\rangle^2 
  \frac{1}{|\hatmass_{11}|}\;,\nonumber  \\
M_2 &\simeq y_2^2 \langle H_u^0\rangle^2
  \left|\frac{\hatmass_{11}}
{\hatmass_{12}^2- \hatmass_{11}\hatmass_{22}} \right|\;,\nonumber  \\
M_3 &\simeq y_3^2 \langle H_u^0\rangle^2|\hatmass^{-1}_{33}|=
y_3^2 \langle H_u^0\rangle^2
  \left|  \frac{\hatmass_{12}^2- \hatmass_{11}\hatmass_{22}}
{{\rm det}\, \hatmass} \right|\;.
\end{align}

Besides, the right-handed mixing matrix reads:
\begin{equation}
V_R= {\rm diag}(e^{i \alpha_1/2}, e^{i (\alpha_2-\alpha_1)/2}, 
e^{i (\alpha_3-\alpha_2)/2})\times W_R\;,
\label{VR}
\end{equation}  
where $\alpha_1={\rm arg}(\hatmass_{11})$,
$\alpha_2={\rm arg}(\hatmass_{11}\hatmass_{22}-\hatmass_{12}^2)$,
$\alpha_3={\rm arg}({\rm det} \hatmass)$ and
\begin{align}
 (W_R)_{12} &\simeq \frac{y_1}{y_2} \frac{\hatmass_{12}}{\hatmass_{11}}\;, & (W_R)_{21} &\simeq -(W_R)_{12}^*\;,\nonumber \\
 (W_R)_{13} &\simeq \frac{y_1}{y_3} \frac{\hatmass_{13}}{\hatmass_{11}}\;, & (W_R)_{31} &\simeq \frac{y_1}{y_3} \frac{\hatmass_{22}^* \hatmass_{13}^*-\hatmass_{12}^*\hatmass_{23}^*}{\hatmass_{12}^{*2}-\hatmass_{11}^*\hatmass_{22}^*}\;, \nonumber \\
 (W_R)_{23} &\simeq \frac{y_2}{y_3} \frac{\hatmass_{12}\hatmass_{13}-\hatmass_{11}\hatmass_{23}}{\hatmass_{12}^2-\hatmass_{11}\hatmass_{22}}\;, & (W_R)_{32} &\simeq -(W_R)_{23}^* \;.
\label{WR}
\end{align}

Thus, the high energy see-saw Lagrangian is parametrized in
terms of the effective neutrino mass matrix, $\mass$, which is in principle
accessible to low energy experiments, the neutrino Yukawa eigenvalues, $D_Y$,
on which we can make the educated guess that they have a hierarchical 
structure, and $V_L$, whose structure is unknown.

\section{Minimal rates for the rare lepton decays}

The supersymmetric see-saw mechanism contains sources of lepton 
flavour violation in the superpotential, encoded in the neutrino
Yukawa matrix $\Yn$, as well as in the soft SUSY breaking 
Lagrangian:
\begin{eqnarray}  
-{\cal L}_{\rm soft}&=&\ 
(\mLs)_{ij} \widetilde L^*_i  \widetilde L_j\ +
(\mes)_{ij} \widetilde e^*_{Ri}  \widetilde e_{Rj}\ + 
(\mns)_{ij} \widetilde \nu^*_{Ri}  \widetilde \nu_{Rj}\ +  
\nonumber \\
&&\ \left(\Ae_{ij} \widetilde e^*_{Ri} H_d \widetilde L_j +
\An_{ij} \widetilde \nu^*_{Ri} H_u \widetilde L_j +
{\rm h.c.}\right)+ 
{\rm etc}\;.
\end{eqnarray}
where $\widetilde L_i$, $\widetilde e_{Ri}$ and $\widetilde \nu_{Ri}$ 
are the supersymmetric partners of the left-handed lepton doublets, 
right-handed charged leptons and right-handed neutrinos, respectively,
$\mLs$, $\mes$ and $\mns$ are their corresponding soft mass matrices squared,
and $\Ae$ and $\An$ are the charged lepton and neutrino soft trilinear
terms. 

The flavour violation in the slepton sector contributes
through one loop diagrams to different flavour violating processes
such as rare muon and tau decays, $K^0_L \rightarrow e^\pm \mu^\mp$ or 
$\mu-e$ conversion in nuclei. Clearly, the minimal rate for 
all those rare processes will arise in a scenario where 
the soft terms are strictly flavour universal at some high energy
scale, $\Lambda$:
\begin{eqnarray}
&&(\mLs)_{ij}=m_L^2 \delta_{ij}, ~~~~
(\mes)_{ij}= m_e^2 \delta_{ij}, ~~~~
(\mns)_{ij}= m_{\nu}^2 \delta_{ij}, \nonumber \\
&&\hspace{1.5cm}(\Ae)_{ij}= \Aen\;{\Ye}_{ij}, ~~~~
(\An)_{ij}= \Ann\;{\Yn}_{ij}\;.
\label{universal}
\end{eqnarray}

If this high energy scale is larger than the 
right-handed neutrino masses, the flavour violation in the
neutrino Yukawa couplings will propagate through radiative effects
to the soft terms~\cite{Borzumati:1986qx}. Hence, even 
under the most conservative assumption for the soft terms from
the point of view of lepton flavour violation, in many
supersymmetric see-saw models some amount of flavour violation 
in the soft SUSY breaking terms is normally expected at low energies .

The off-diagonal elements of the soft SUSY breaking terms read
at low energies, in the leading-log approximation,~\footnote{Note 
that the result for $(\Ae)_{ij}$ differs
from the one usually quoted in the literature, which is
proportional $(2 A_\nu+A_e)/(16\pi^2)$. The reason is that
quantum corrections due to right-handed neutrinos also
induce off-diagonal terms in the charged lepton Yukawa couplings.
Hence, at low energies it is necessary to redefine the charged
lepton basis in order to bring the charged lepton Yukawa coupling
to its diagonal form. This introduces new sources of flavour
violation in the soft terms, which are negligible in $\mLs$
but not in $\Ae$. Indeed, these new sources of flavour violation
have the effect of removing the dependence in $A_e$ in the
off-diagonal trilinear terms $(\Ae)_{ij}$~\cite{Ibarra:2008uv}.}
\begin{eqnarray}
\left(\mLs\right)_{ij} & \simeq & -\frac{1}{8\pi^2}
(m_L^2+m_\nu^2+\mHus+|\Ann|^2) P_{ij} \ , \nonumber\\
\left(\mes\right)_{ij} & \simeq & 0\ , \nonumber\\  
(\Ae)_{ij}&\simeq &   \frac{-1}{8\pi^2} \Ann \Ye_{ii} P_{ij}\;, 
\label{softafterRG} 
\end{eqnarray}
where $i\neq j$ and
\begin{equation}
P_{ij} =\sum_k {{\bf Y}_\nu^*}_{ki} \log \left(\frac{\Lambda}{M_k}\right) \Yn_{kj}\;.
\label{P-def}
\end{equation}

The size of the off-diagonal soft terms depends crucially
on the flavour structure of the neutrino Yukawa couplings
and on the scale of the cut-off, $\Lambda$, which can be identified 
with the mass of the messenger particles which transmit
supersymmetry breaking from the hidden sector to the
observable sector. We will show in this paper that 
if thermal leptogenesis is the correct mechanism to generate
the baryon asymmetry, a non-vanishing rate for the
rare decays will be {\it necessarily} generated,
unless artificial cancellations among different terms
are taking place. 

In the simplest version of the leptogenesis mechanism, 
the lightest right-handed neutrino is produced by thermal
scatterings in the primeval plasma. Subsequently, the
out of equilibrium decays of the right-handed neutrinos 
generate a lepton asymmetry, which is eventually converted
by sphaleron processes into a baryon asymmetry. In order
to produce the observed baryon asymmetry by this
mechanism the reheating temperature of the Universe
has to be larger than $\sim 10^9$ GeV~\cite{Davidson:2002qv,Buchmuller:2004nz}.
At these very high temperatures gravitino thermal production is very 
efficient, therefore, in order to avoid overclosure of the Universe
the gravitino mass has to be larger than $m_{3/2}\gsim 5$ GeV
\cite{Bolz:2000fu,Pradler:2006qh}, which implies a rather 
large scale for the cut-off.

To show this, we recall that the gravitino mass is defined as
\begin{equation}
m_{3/2}=\frac{|F|}{\sqrt{3}M_P}\;,
\label{gravitino-mass}
\end{equation}
where $M_P=2.4 \times 10^{18}$ GeV is the reduced Planck mass
and $\sqrt{|F|}$ is the scale of spontaneous supersymmetry breaking.
On the other hand, SUSY breaking is transmitted to the observable
sector by messenger particles with mass $M_{\rm mes}$, inducing
soft masses which approximately read:
\begin{equation}
m^2_{\rm soft} \sim c \frac{|F|^2}{M_{\rm mes}^2}\;,
\label{soft-mass}
\end{equation}
where $c\sim 10^{-4}-1$ is a constant which depends on the details of the 
mediation mechanism. From 
Eqs.~(\ref{gravitino-mass},\ref{soft-mass}) it follows that
\begin{equation}
M_{\rm mes}\sim \sqrt{3c}\frac{m_{3/2}}{m_{\rm soft}} M_P \;.
\end{equation}
Therefore, the constraint on the gravitino mass
from the requirement of successful leptogenesis, $m_{3/2}\gsim 5$ GeV,
 and the assumption 
that the soft masses are ${\cal O}(1{\rm TeV})$ imply that the messenger
scale has to be larger than $10^{14}-10^{16}$ GeV. This
large scale for the cut-off suggests that at least one right-handed
neutrino is coupled below the mediation scale and thus will 
contribute to the generation of off-diagonal soft terms 
via quantum effects~\cite{Tobe:2003nx}\footnote{ This bound
on the messenger scale could be circumvented if the gravitino is ultralight
so is in thermal equilibrium in the early Universe, namely
$m_{3/2}\lsim 16$ eV, which corresponds to $M_{\rm mes}\lsim 260$ TeV
\cite{Viel:2005qj}. This scenario requires, though, an
extension of the Minimal Supersymmetric Standard Model in order
to account for the cold dark matter of the Universe, since neither
the gravitino nor the lightest neutralino are any 
longer good dark matter candidates.}.

Indeed, the experimental fact that the ratio of the atmospheric mass 
splitting to the solar mass splitting is relatively mild, $\sqrt{\Delta m^2_{\rm atm}/\Delta m^2_{\rm sol}}\sim 6$, supports this conclusion.
As discussed in \cite{Casas:2006hf}, when the neutrino Yukawa eigenvalues
are hierarchical, a degenerate spectrum of right-handed neutrinos
cannot reproduce the observed mild neutrino mass hierarchy without
a certain fine tuning. This is not the case, though, for a hierarchical
spectrum of right-handed neutrinos, which could naturally explain
the neutrino mass hierarchy for certain choices of the matrix $V_R$
without tunings. Therefore, even assuming that the heaviest right-handed
neutrino mass is around the Planck scale, in view of the large hierarchy
necessary to accommodate the ratio of the solar and atmospheric mass splittings
without fine-tuning, it is plausible that at least the lightest right-handed
neutrino will have a mass smaller than $10^{14}-10^{16}$ and hence
will contribute to the radiative generation of off-diagonal terms 
in the leptonic soft terms.

The second necessary requirement to generate radiatively flavour
violation in the soft SUSY breaking terms is a non trivial
structure in the neutrino Yukawa couplings, encoded in the
matrix $P$, Eq.~(\ref{P-def}). This equation can 
be conveniently rewritten as
\begin{equation}
P_{ij}=({\bf Y}_\nu^\dagger \Yn)_{ij}\log\left(\frac{\Lambda}{M_3}\right)+
{\bf Y}^*_{2i} \Yn_{2j} \log\left(\frac{M_3}{M_2}\right)+
{\bf Y}^*_{1i} \Yn_{1j} \log\left(\frac{M_3}{M_1}\right)\;.
\label{P-terms}
\end{equation}

For generic neutrino Yukawa couplings, this expression is
dominated by the first term, which corresponds
to the widely used approximation of decoupling all 
the right-handed neutrinos altogether at the scale $M_3$.
However, it is conceivable that the matrix ${\bf Y}_\nu^\dagger \Yn$ 
could be exactly diagonal. If this is the case, the leading
contributions to the off diagonal elements of $P$ are
determined by the subdominant terms proportional to $\Yn_{2i}$ and $\Yn_{1i}$.
In this scenario, that as we will see is consistent with
present neutrino experiments, the mixing in the left-handed
sector is trivial, namely $V_L=1$. However, in order to 
generate mixing in the effective neutrino mass matrix,
there must exist mixing in the right-handed sector, $V_R \neq 1$.
Therefore, even in this extreme scenario, a non vanishing rate
for the rare decays will always be generated through the
subdominant terms $\Yn_{2i}=(V_R)_{2i} y_i$ and $\Yn_{1i}=(V_R)_{1i} y_i$,
unless different terms cancel each other.
More concretely, using Eqs.~(\ref{VR},\ref{WR}) it follows that in the
scenario with $V_L=1$, when $\Lambda>M_3$,
\begin{align}
\label{Ps}
 P_{12} &\simeq
y_1^2 \frac{\mass_{12}}{\mass_{11}} \log  \frac{M_2}{M_1}\;, \nonumber  \\
P_{13} &\simeq
y_1^2 \Big[\frac{\mass_{12}\mass_{23}-\mass_{13}\mass_{22}}{\mass^2_{12}-\mass_{11}\mass_{22}} \log  \frac{M_3}{M_2} 
+\frac{\mass_{13}}{\mass_{11}} \log \frac{M_2}{M_1}\Big]\;,\nonumber  \\
 P_{23} &\simeq
y_2^2 \frac{\mass_{12}\mass_{13}-\mass_{11}\mass_{23}}{\mass_{12}^2-\mass_{11}\mass_{22}} \log \frac{M_3}{M_2} \;,
\end{align}
on the other hand, when $M_3>\Lambda>M_2$, the expressions are
identical with the substitution $M_3\rightarrow \Lambda$.
Lastly, when $M_2>\Lambda>M_1$,
\begin{align}
 P_{12} &\simeq
y_1^2 \frac{\mass_{12}}{\mass_{11}} \log  \frac{\Lambda}{M_1}\;, \nonumber  \\
P_{13} &\simeq 
y_1^2\frac{\mass_{13}}{\mass_{11}} \log \frac{\Lambda}{M_1}\;,\nonumber \\
 P_{23} &\simeq
y_1^2 \frac{\mass^*_{12} \mass_{13}}{|\mass_{11}|^2} 
\log \frac{\Lambda}{M_1} \;,
\end{align}
where the right-handed neutrino masses are given in Eq.~(\ref{RH-masses}).

The off diagonal elements of the matrix $P$ induce through quantum
corrections flavour violation in the soft mass matrices,
 Eq.~(\ref{softafterRG}),
which in turn induce a non-vanishing rate for the rare lepton
decays, which approximately reads:
\begin{eqnarray}
{\rm BR}(\l_i\rightarrow \l_j \gamma) &\simeq& \frac{\alpha^3}{G_F^2}
\frac{|(\mLs)_{ij}|^2}{m_S^8}\tan^2\beta\, {\rm BR}(\l_i\rightarrow \l_j \nu_i \bar \nu_j)\;,
\label{BR}
\end{eqnarray}
where ${\rm BR}(\mu\rightarrow e \nu_\mu \bar \nu_e)\simeq 1$,
 ${\rm BR}(\tau\rightarrow \mu \nu_\tau \bar \nu_\mu)\simeq 0.17$,
${\rm BR}(\tau\rightarrow e \nu_\tau \bar \nu_e)\simeq 0.18$, and
$m_S$ is a mass scale of the order of typical SUSY masses.

Among the scenarios compatible with the present neutrino 
experiments and thermal leptogenesis, the one presented here,
with flavour universal soft terms at some cut-off scale
and no flavour mixing in the left-handed sector, 
corresponds to the worst case for the detection of
the rare decays or, conversely, to the scenario yielding
the minimal rate for the rare decays. In any other scenario
there will be additional sources of flavour violation, either
in the soft terms at the cut-off scale or in the left-handed
mixing matrix $V_L$, thus yielding a larger rate for the rare
decays, unless different terms cancel each other.

In what follows, let us illustrate our results calculating
the minimal rates for the rare decays in the Constrained MSSM, 
which is defined at the Grand Unification scale
by just five parameters: the
universal scalar mass ($m_0$), gaugino mass ($M_{1/2}$) and
trilinear term ($A_0$), $\tan\beta$ and the sign of $\mu$.
As neutrino parameters, we will assume a hierarchical mass
spectrum (which is the most plausible possibility under
the assumption of hierarchical Yukawa eigenvalues~\cite{Casas:2006hf})
and a neutrino mixing matrix approximately 
tri-bimaximal~\cite{Harrison:2002er}:
\begin{equation}
U\approx \begin{pmatrix} \sqrt{\frac{2}{3}} & \sqrt{\frac{1}{3}} & 0 \\
-\sqrt{\frac{1}{6}} & \sqrt{\frac{1}{3}} & -\sqrt{\frac{1}{2}} \\
-\sqrt{\frac{1}{6}} & \sqrt{\frac{1}{3}} & \sqrt{\frac{1}{2}}
\end{pmatrix} 
\times {\rm diag}(e^{i \phi/2},e^{i \phi^\prime/2},1)\;
\end{equation}
(assuming a non-vanishing $|U_{13}|$ will not change our conclusions).
Then, the minimal rates for the rare decays can be straightforwardly
computed using Eqs.~(\ref{softafterRG},\ref{Ps},\ref{BR}), yielding
\begin{align}
\label{minimal-rates}
{\rm BR}(\mu \rightarrow e \gamma) &\gsim \frac{\alpha^3}{G_F^2}
\left(\frac{3 m^2_0 +|A_0|^2}{8\pi^2 m_S^4}\right)^2
 y^4_1 \log^2\frac{M_2}{M_1}  \tan^2\beta \;,\nonumber  \\
{\rm BR}(\tau \rightarrow e \gamma) &\gsim \frac{\alpha^3}{G_F^2}
\left(\frac{3 m^2_0 +|A_0|^2}{8\pi^2 m_S^4}\right)^2
 y^4_1 \left(2 \log\frac{M_3}{M_2}+\log\frac{M_2}{M_1}\right)^2
\tan^2\beta 
\;{\rm BR}(\tau \rightarrow e \nu_\tau \bar \nu_\mu)\;, \nonumber \\
{\rm BR}(\tau \rightarrow \mu \gamma) &\gsim \frac{\alpha^3}{G_F^2}
\left(\frac{3 m^2_0 +|A_0|^2}{8\pi^2 m_S^4}\right)^2
 y^4_2 \log^2\frac{M_3}{M_2} \tan^2\beta 
\;{\rm BR}(\tau \rightarrow \mu \nu_\tau \bar \nu_e) \;,
\end{align}
which strongly depend on the size of the Yukawa eigenvalues 
and only logarithmically on the hierarchy of right-handed masses,
or alternatively, on the hierarchy of the Yukawa eigenvalues,
through 
\begin{align}
\frac{M_3}{M_2}&\simeq \frac{y^2_3}{y^2_2} \frac{m_3}{12 m_1}\;, \nonumber  \\
\frac{M_2}{M_1}&\simeq \frac{y^2_2}{y^2_1} \frac{2 m_2}{3 m_3} \;.
\end{align}

For given neutrino Yukawa eigenvalues one can estimate using
 Eq.~(\ref{minimal-rates}) a lower bound on the rates of the 
rare lepton decays. Conversely, one can derive constraints on 
the parameters of the high-energy Lagrangian $y_1$ and $y_2$ 
from the present bounds on the rare lepton decays, 
${\rm BR}(\mu \rightarrow e \gamma)\leq 1.2\times 10^{-11}$~\cite{Brooks:1999pu},
${\rm BR}(\tau \rightarrow \mu \gamma)\leq 4.5\times 10^{-8}$~\cite{Hayasaka:2007vc},
${\rm BR}(\tau \rightarrow e \gamma)\leq 1.1\times 10^{-7}$~\cite{Aubert:2005wa}.
The most stringent constraints on $y_1$ and $y_2$ stem from the
non-observation of the processes $\mu \rightarrow e \gamma$ and 
$\tau \rightarrow \mu \gamma$, respectively, and read:
\begin{align}
\label{bounds-Yukawas}
y_1&\lsim 4\times 10^{-2} 
\left(\frac{{\rm BR}(\mu \rightarrow e \gamma)}{1.2\times 10^{-11}}\right)^{1/4}
\left(\frac{m_S}{200\, {\rm GeV}}\right)
\left(\frac{\tan\beta}{10}\right)^{-1/2}\;,\nonumber  \\
y_2&\lsim 0.5
\left(\frac{{\rm BR}(\tau \rightarrow \mu \gamma)}{4.5\times 10^{-8}}\right)^{1/4}
\left(\frac{m_S}{200\, {\rm GeV}}\right)
\left(\frac{\tan\beta}{10}\right)^{-1/2} \;,
\end{align}
where we have conservatively assumed $M_3:M_2:M_1=100:10:1$ and 
$m_0\sim A_0 \sim m_S$. Note that the bound on $y_2$ only applies
when $\Lambda> M_2$.

This numerical estimate is confirmed by our numerical analysis
of two typical points in the CMSSM parameter space.
We have analyzed the SPS1a and SPS1b benchmark points~\cite{Allanach:2002nj}, 
which correspond to typical CMSSM points with intermediate and 
relatively high values of $\tan\beta$, respectively. For these two benchmark
points we find approximately the same result:
\begin{align}
y_1&\lsim 6\times 10^{-2} 
\left(\frac{{\rm BR}(\mu \rightarrow e \gamma)}{1.2\times 10^{-11}}\right)^{1/4}\;,\nonumber \\
y_2&\lsim 0.8
\left(\frac{{\rm BR}(\tau \rightarrow \mu \gamma)}{4.5\times 10^{-8}}\right)^{1/4}\;,
\end{align}
which agrees with our general expectation, Eq.~(\ref{bounds-Yukawas}).

Alternatively, these bounds could be expressed in terms of
the SUSY contribution to the muon $g-2$, which depends on the same
combination of SUSY masses and $\tan\beta$~\cite{g-2},
\begin{equation}
\delta a_\mu^{\rm SUSY}\simeq \frac{5 g_2^2}{192\pi^2}
\frac{m_\mu^2}{m^2_S}\tan\beta\;,
\end{equation}
yielding
\begin{align}
y_1&\lsim 6\times10^{-2} 
\left(\frac{{\rm BR}(\mu \rightarrow e \gamma)}{1.2\times 10^{-11}}\right)^{1/4}
\left(\frac{\delta a^{\rm SUSY}_\mu}{10^{-9}}\right)^{-1/2}
\;,\nonumber \\
y_2&\lsim 0.8
\left(\frac{{\rm BR}(\tau \rightarrow \mu \gamma)}{4.5\times 10^{-8}}\right)^{1/4}
\left(\frac{\delta a^{\rm SUSY}_\mu}{10^{-9}}\right)^{-1/2}\;,
\end{align}

These bounds demonstrate that it is possible to obtain information on
the high energy see-saw parameters from low energy observations
(namely the bounds on the rates of the rare decays, neutrino masses
and mixing angles and supersymmetric parameters), under very
general and well motivated assumptions about the high energy theory,
such as the absence of cancellations, hierarchical neutrino Yukawa
eigenvalues and a large mediation scale (as suggested by thermal
leptogenesis). 

The resulting bound on $y_2$ is rather weak and lacks any practical
interest. On the other hand, the bound on $y_1$ is fairly stringent
(it corresponds to a Dirac neutrino mass of 7 GeV) and will be improved
in the near future by a factor of three if the MEG experiment 
at PSI reaches the projected
sensitivity ${\rm BR}(\mu\rightarrow e \gamma)\sim 10^{-13}$~\cite{MEG}
without observing a positive signal. Furthermore, the bound on $y_1$
has important implications for leptogenesis, which will be discussed
in the next section.

A similar rationale could be applied to calculate the minimal
value of the leptonic electric dipole moments (EDMs). Following
the analysis in \cite{Masina:2003wt}, we estimate that in the 
worst case scenario  for the detection of EDMs, again when the 
soft terms are flavour universal at the cut-off scale and when $V_L=1$,
the electron EDM reads:
\begin{equation}
 d_e \sim e \frac{\alpha}{\pi} \frac{m_e}{m_S^2} \left( \frac{1}{16 \pi^2}\right)^2 y_1^4 \; {\rm Im} \left[
 \frac{\mass_{12}\mass_{13}}{\mass_{11}^2} \frac{\mass_{12}\mass_{13}-\mass_{11}\mass_{23}}{\mass_{12}^2-\mass_{11}\mass_{22}}\right]\log \frac{M_2}{M_1} \log \frac{M_3}{M_2}\;,
\end{equation}
where we have assumed $A_0 \sim \mu \sim M_1 \sim m_S$. Then, when
the neutrino mass matrix has an approximate tri-bimaximal form
and allowing a non-vanishing value for the 13 element of the leptonic
mixing matrix,   $U_{13}=\sin\theta_{13} e^{-i\delta}$, we find 
the following lower bound on the electron EDM:
\begin{equation}
 |d_e| \gsim e \frac{\alpha}{\pi} \frac{m_e}{m_S^2} \left( \frac{1}{16 \pi^2}\right)^2 y_1^4 \, \left|
2\sqrt{2} \sin\theta_{13} \sin\delta+
6 \frac{m_1}{m_2} \sin (\phi^\prime-\phi)
\right|\log \frac{M_2}{M_1} \log \frac{M_3}{M_2}\;,
\end{equation}
which can even be exactly zero if there is no CP violation at low energies
or when both $m_1$ and $\sin\theta_{13}$ simultaneously vanish.
Assuming generic CP violating phases and 
$\sin\theta_{13}=0.2$, which corresponds to the present upper 
bound at the 2$\sigma$ level~\cite{Schwetz:2008er}, the following
lower bound holds:
\begin{equation}
 |d_e| \gsim
 7\times 10^{-29}\,y_1^4\, {\rm e}\, {\rm cm} \left( \frac{m_S}{200 {\rm GeV}}\right)^{-2}\;.
\end{equation}
Finally, from the present experimental bound on the electron EDM,
$|d_e|<10^{-27}\, {\rm e}\, {\rm cm}$~\cite{Regan:2002ta}, 
we obtain the constraint
on the smallest Yukawa eigenvalue $y_1\lsim 2$ 
for $m_S=200$ GeV, which is much weaker than the bound we derived in
Eq.~(\ref{bounds-Yukawas}) from the non-observation of the process 
$\mu\rightarrow e\gamma$. More importantly, the constraint on $y_1$ 
from the electron EDM relies on assumptions about
the size of the CP violating phases and $\sin\theta_{13}$, which are
currently unknown. We find that 
even if future experiments determine
that the CP violating phases and $\sin\theta_{13}$ are sizable,
the best current proposal to improve the experimental sensitivity
to the electron EDM will not provide bounds on $y_1$ competitive
to the bounds stemming from the non-observation of  $\mu\rightarrow e \gamma$.
Namely, an improvement of sensitivity down to the level 
$d_e\sim 10^{-35}\,{\rm e}\, {\rm cm}$~\cite{Lamoreaux:2001hb}, 
would translate into $y_1\lsim 0.02$, again for $m_S=200$ GeV,
which is comparable to the bound attainable by the MEG experiment
at PSI, provided no positive signal is found.

\section{Implications for leptogenesis}

The baryon asymmetry generated through the leptogenesis mechanism
depends, under the assumption of hierarchical neutrinos,
essentially on two parameters: the lightest right-handed neutrino mass, 
$M_1$, and an effective neutrino mass $\mtilde1$~\cite{lepto-review}, 
defined as
\begin{equation}
\mtilde1 = \frac{( {\bf Y}^{}_\nu{\bf Y}^\dagger_\nu)_{11}}{M_1} 
\langle H_u^0\rangle^2\;,
\label{mtilde1} 
\end{equation}
which measures the strength of the coupling of the lightest right-handed
neutrino to the thermal bath.

In scenarios where the neutrino Yukawa couplings are hierarchical
the lightest right-handed neutrino mass reads,
for generic values of the matrix $V_L$,  
\begin{equation}
\label{M1}
M_1 \simeq \frac{y_1^2 \langle H_u^0\rangle^2}{|\hatmass_{11}|}\;,
\end{equation}
where $|\hatmass_{11}|= |(V^T_L \mass V_L)_{11}|
=|\sum_k (U^\dagger V_L)^2_{k1} m_k|$.
Strictly speaking, $|\hatmass_{11}|$ can range between 0 and $m_3$.
Nevertheless, for generic values of the matrix $V_L$ 
the most natural range for $|\hatmass_{11}|$ is 
$\sqrt{\Delta m^2_{\rm sol}}\lsim |\hatmass_{11}|\lsim 
\sqrt{\Delta m^2_{\rm atm}}$. 
The only exception corresponds
to the case when $(U^\dagger V_L)_{k1}\simeq \delta_{k1}$,
which can arise in specific models and which leads to 
$|\hatmass_{11}|\ll\sqrt{\Delta m^2_{\rm sol}}$ without cancellations.
This special case will be discussed at the end of this section.
In our numerical analysis we will take
for the solar and atmospheric mass splittings
the central values  of the global fit to neutrino data~\cite{Schwetz:2008er},
$\Delta m^2_{\rm sol}=7.65\times 10^{-5}\,{\rm eV}^2$,
$\Delta m^2_{\rm atm}=2.40\times 10^{-3}\,{\rm eV}^2$.

In the generic case $\sqrt{\Delta m^2_{\rm sol}}\lsim |\hatmass_{11}|\lsim 
\sqrt{\Delta m^2_{\rm atm}}$. Then, from  Eq.~(\ref{M1}) it follows 
a natural range for $M_1$ as a function of $y_1$. More importantly, 
the lightest neutrino Yukawa eigenvalue, $y_1$, 
is bounded from above by the non-observation of the process 
$\mu\rightarrow e\gamma$, through $|P_{12}|\gsim y^2_1 \log M_2/M_1$,
 {\it cf.}  Eq.~(\ref{Ps}).  
Therefore, in a supersymmetric scenario with hierarchical 
neutrino Yukawa couplings, 
the following {\it upper} bound on the lightest right-handed neutrino mass
holds for generic values of the matrix $V_L$:
\begin{equation}
M_1 \lsim |P_{12}| 
      \frac{\langle H_u^0\rangle^2}{\sqrt{\Delta m^2_{\rm sol}}}
\log^{-1} \frac{M_2}{M_1}\;,
\label{M1fromP12}
\end{equation}
which numerically reads
\begin{equation}
M_1 \lsim 5\times 10^{12}{\rm GeV} 
\left(\frac{{\rm BR}(\mu \rightarrow e \gamma)}
{1.2\times 10^{-11}}\right)^{1/2}
\left(\frac{m_S}{200\, {\rm GeV}}\right)^2
\left(\frac{\tan\beta}{10}\right)^{-1} \;.
\label{M1fromBR}
\end{equation}
This upper bound should be compared with the {\it lower} 
bound on the right-handed neutrino mass $M_1\gsim 10^9$ GeV,
leaving an allowed window of four orders of magnitude
for the lightest right-handed neutrino mass.
Alternatively, Eq.~(\ref{M1fromBR}) could be rewritten
as a lower bound on the rate for $\mu \rightarrow e \gamma$
as a function of the lightest right-handed neutrino mass,
\begin{equation}
{\rm BR}(\mu \rightarrow e \gamma)\gsim 5\times 10^{-19} 
\left(\frac{M_1}{10^9 \,{\rm GeV} }\right)^2
\left(\frac{m_S}{200\, {\rm GeV}}\right)^{-4}
\left(\frac{\tan\beta}{10}\right)^{2} \;.
\label{BRfromM1}
\end{equation}
Thus, exploring the allowed window of the thermal leptogenesis
scenario requires an improvement in sensitivity to the process
${\rm BR}(\mu \rightarrow e \gamma)$ of approximately eight orders of
magnitude, which unfortunately does not seem feasible in the short or mid
term. It is remarkable, though, that if supersymmetry is discovered
at the LHC, the scenario of thermal leptogenesis with hierarchical 
neutrino Yukawa couplings and hierarchical right-handed masses
could be tested using just low energy experiments.

Interestingly, following our premise of the absence of cancellations,
the allowed mass window for leptogenesis can be further narrowed down. In
the scenario with hierarchical neutrino Yukawa eigenvalues,
the effective neutrino mass $\mtilde1$ reads, following 
Eqs.~(\ref{VR},\ref{WR}):
\begin{equation}
\mtilde1 \simeq \frac{|\hatmass_{11}|^2+|\hatmass_{12}|^2+|\hatmass_{13}|^2}{|\hatmass_{11}|}\;,
\label{mtilde-hierarchical}
\end{equation}
which ranges between $m_1\leq \mtilde1 < \infty$ for the Yukawa couplings 
consistent with the low energy neutrino experiments~\cite{Davidson:2002qv}.
Using $|\hatmass_{1i}|=|\sum_k (U^\dagger V_L)_{k1}(U^\dagger V_L)_{ki} m_k|$
it follows that  the lower limit, $\mtilde1=m_1$, could be reached 
when $(V_L)_{k1}=U_{k1}$, which corresponds to the special case for
the matrix $V_L$ which will be discussed at the end of this section.
On the other hand, the upper limit, $\mtilde1\rightarrow \infty$ 
is reached when $\sum_k (U^\dagger V_L)^2_{k1} m_k =0$, which requires 
a cancellation among terms and is thus implausible. 
Therefore, in the generic case $\sqrt{\Delta m^2_{\rm sol}}\lsim |\hatmass_{11}|\lsim \sqrt{\Delta m^2_{\rm atm}}$, one expects a natural window
for the effective neutrino mass
$\sqrt{\Delta m^2_{\rm sol}}\lsim \mtilde1 \lsim
\sqrt{\Delta m^2_{\rm atm}}$,
which corresponds to the strong washout regime.
On the other hand,
from the analysis in \cite{Blanchet:2006be,JosseMichaux:2007zj} it follows that 
an effective neutrino mass  $\mtilde1\gsim \sqrt{\Delta m^2_{\rm sol}}$ implies
a lower bound on the right-handed neutrino mass  $M_1\gsim 3\times 10^{9}$
GeV, which in turn implies, following Eq.~(\ref{BRfromM1}),  the lower 
bound on the rare muon decay
${\rm BR}(\mu \rightarrow e \gamma)\gsim 5\times 10^{-18}$
for typical SUSY parameters. Therefore, closing the natural window 
for leptogenesis requires, for generic neutrino parameters, 
an improvement in sensitivity to the process 
$\mu\rightarrow e\gamma$ of six orders of magnitude.

The required sensitivity is unfortunately below,
although not far from, the sensitivity of the projected
neutrino factory, where the high beam intensity may allow the observation of
one $\mu\rightarrow e\gamma$ event if the branching ratio is $10^{-16}$.
One should note, however, that the observation of this single event over the 
accidental background would require detector resolutions which are not 
currently available, and new technologies or new experimental ideas 
should be developed~\cite{Bandyopadhyay:2007kx}. On the other hand,
the PRISM/PRIME experiment at J-PARC aims 
to achieve a single event sensitivity to 
the process $\mu\; {\rm Ti}\rightarrow e\; {\rm Ti}$ at the level of
$10^{-18}$~\cite{PRISM}. This is equivalent to a sensitivity to the process 
$\mu \rightarrow e \gamma$ at the level 
of $\sim 2\times 10^{-16}$.~\footnote{When
the photon penguin diagram dominates the $\mu-e$ conversion in Ti, the 
conversion rate is approximately a factor $5\times 10^{-3}$ smaller than
the branching ratio of  $\mu \rightarrow e \gamma$.}
Thus, if the LHC determines that $\tan\beta$ is large,
the non-observation of muon
flavour violation at PRISM/PRIME could rule out, for generic
neutrino parameters and barring cancellations, the thermal 
leptogenesis scenario based on the decay of the lightest right-handed
neutrino. If, on the contrary, $\tan\beta$ takes moderate values, 
it would be necessary a
further improvement in sensitivity by more than one order of magnitude
to close the leptogenesis window for $M_1$.

For large values of $\mtilde1$, the upper bound on the lightest 
right-handed neutrino mass can be improved. From 
Eqs.~(\ref{M1},\ref{mtilde-hierarchical}) it follows that:
\begin{equation}
\mtilde1\simeq\frac{y^2_1 \langle H_u^0\rangle^2}{M_1}
\left[1+ \left|\frac{\hatmass_{12}}{\hatmass_{11}}\right|^2+
\left|\frac{\hatmass_{13}}{\hatmass_{11}}\right|^2\right]\;.
\end{equation}
Barring cancellations one expects in general
$|\hatmass_{11}|\sim |\hatmass_{12}|\sim |\hatmass_{13}|$.
Then, using the upper bound on the lightest Yukawa coupling, 
Eq.~(\ref{bounds-Yukawas}), we obtain:
\begin{equation}
M_1 \lsim 10^{13} {\rm GeV}\left(\frac{\mtilde1}{9\times 10^{-3}\,{\rm eV}}\right)^{-1}
\left(\frac{{\rm BR}(\mu \rightarrow e \gamma)}{1.2\times 10^{-11}}\right)^{1/2}
\left(\frac{m_S}{200\, {\rm GeV}}\right)^2
\left(\frac{\tan\beta}{10}\right)^{-1} \;.
\label{M1-mtilde}
\end{equation}

From Eq.~\eqref{M1fromBR} it is apparent 
that already large portions of the parameter space for $M_1$ are excluded.
In particular this bound suggests that flavour effects should be always taken into
account in leptogenesis. More concretely,
the lepton asymmetry is mostly generated at the temperature $T_B$,
defined through~\cite{Buchmuller:2004nz}
\begin{equation}
\frac{M_1}{T_B}\simeq1+\frac{1}{2}\log\left(1+\frac{\pi K^2}{1024}
\left[\log\left(\frac{3125\pi K^2}{1024}\right)\right]^5\right)\;,
\end{equation}
where $K=\mtilde1/m_*$ and $m_*\simeq 10^{-3}$ eV. For $T>T_B$
the asymmetry produced is essentially erased, while for $T<T_B$,
washout is negligible. In the strong washout regime $K\gg1$,
which in turn implies $T_B\lsim M_1 \lsim 5\times 10^{12}$ GeV.
In this range of temperatures the tau Yukawa coupling
is in equilibrium and thus flavour effects can be relevant.
For a hierarchical left-handed neutrino spectrum
there are two possibly relevant flavour effects,
lowering the bound on $M_1$ in the strong washout regime \cite{Abada:2006fw}.
If the lightest right-handed neutrino decays into different flavours,
the washout in each flavour is not determined by $\mtilde1$, but instead by
\begin{equation}
 \mtildea = \frac{ |{\bf Y}^{}_\nu|_{1\alpha}^2}{M_1} 
\langle H_u^0\rangle^2\;.
\end{equation}
As $\mtilde1 = \sum_\alpha \mtildea$,
the washout in each flavour is smaller than in the unflavoured case.
In the approximation of flavours being CP eigenstates,
the flavoured CP asymmetries are proportional to $\mtildea$ and
the effect is maximized for equal $\mtildea$ in all flavours $\alpha$.
This can lead to a relaxation of the lower bound on $M_1$ by a factor 2--3,
depending on the number of families 
which have charged lepton Yukawa interactions in equilibrium with
the thermal plasma.
Secondly, for some specific neutrino textures it may occur
that the CP asymmetry is sizable in one flavour, but the asymmetry is
only weakly washed out, $\mtildea\sim m_*$. Using 
Eqs.~(\ref{VR},\ref{WR}) it follows that
\begin{equation}
 \mtildea \simeq \frac{|V_L^T {\cal M}|_{1\alpha}^2}{|\hatmass_{11}|}\;,
\end{equation}
from where 
it is apparent that this possibility requires $V_L$ with sizable
off-diagonal entries (unless the low energy phases and $|U_{13}|$ 
take very special values), thus leading to an enhancement of
${\rm BR}(\mu\rightarrow e\gamma)$. Since we are
interested in scenarios yielding the minimal rate for $\mu\rightarrow e\gamma$
we will not consider this possibility.

In weak washout the lower bound on $M_1$ is not relaxed by flavour effects
and especially the absolute lower bound $M_1\gsim 10^9$ GeV is not affected \cite{Blanchet:2006be,JosseMichaux:2007zj}.

\begin{figure}
\begin{center}
\begin{tabular}{c}
\epsfig{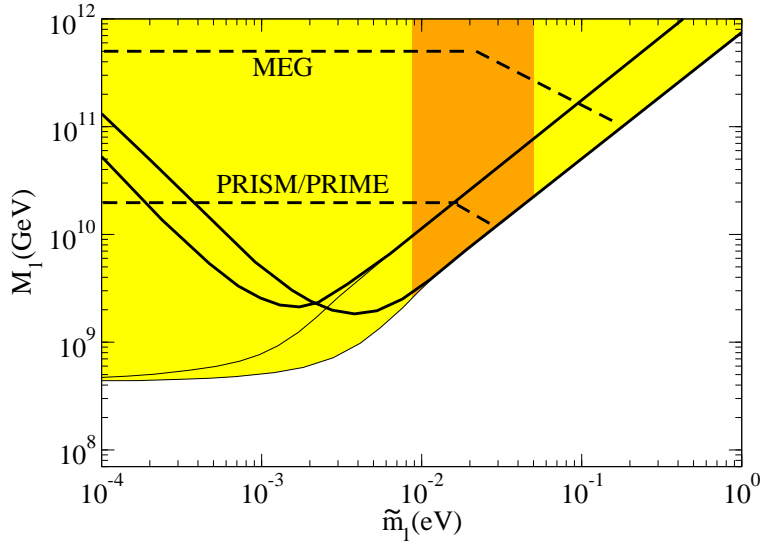} 
\end{tabular}
\end{center}
\caption
{\small Allowed parameter space of thermal leptogenesis (in yellow,
adopted from \cite{Blanchet:2006be}), including the constraints
on the relevant parameters which stem from the non-observation
of the process $\mu\rightarrow e\gamma$, under the assumption
of hierarchical neutrino Yukawa eigenvalues and barring cancellations.
The orange region corresponds to the range of $\mtilde1$ for 
generic neutrino parameters. In this plot it is assumed $m_S\simeq 200$
GeV and $\tan\beta\simeq 10$.
}
\label{figure1}
\end{figure}


We show in Fig.~\ref{figure1} the impact of the bounds on
the lightest right-handed neutrino mass stemming from the non-observation
of the process $\mu\rightarrow e\gamma$, Eqs.~(\ref{M1fromBR},\ref{M1-mtilde}),
on the parameter space of thermal leptogenesis, spanned by  $\mtilde1$ 
and $M_1$. The yellow region corresponds to the allowed region found 
by Blanchet and di Bari, and shown in 
Fig.~1 of ~\cite{Blanchet:2006be}.
 The thick solid lines 
encompass the allowed region assuming zero initial abundance
of right-handed neutrinos, while the thin solid lines, the
allowed region assuming thermal initial abundance.
For each case we show the lower bound on $M_1$ for two
scenarios. The left plot corresponds to the ``alignment'' scenario,
where the final asymmetry is dominated by one flavor, and which amounts to
neglecting flavour effects in leptogenesis. The right plot
corresponds to the ``democratic'' scenario, where
$\mtildea=\mtilde1 /3$, and which illustrates how flavour effects can
relax the lower bound on $M_1$.
On the other hand, the 
orange region corresponds to the range of $\mtilde1$ for 
generic neutrino parameters,
$\sqrt{\Delta m^2_{\rm sol}}\lsim \mtilde1 \lsim \sqrt{\Delta m^2_{\rm atm}}$.
We show as thick dashed lines the most stringent upper bound on 
$M_1$
for the projected sensitivity by the MEG experiment at PSI,
${\rm BR}(\mu \rightarrow e \gamma)\sim 10^{-13}$,
corresponding to $M_1\lsim 5\times 10^{11}$ GeV,
and the projected sensitivity by the PRISM/PRIME experiment at J-PARC,
${\rm R} (\mu\; {\rm Ti}\rightarrow e\; {\rm Ti}) \sim 10^{-18}$
which corresponds to $M_1\lsim 2\times 10^{10}$ GeV.
It should be stressed at this point that
the main uncertainty in our calculation does not stem from the calculation 
of the baryon asymmetry, but from our present ignorance of the SUSY 
parameters, which can change considerably our numerical estimate of 
${\rm BR}(\mu \rightarrow e \gamma)$.

Lastly, we would like to discuss the case with $|\hatmass_{11}|\ll 
\sqrt{\Delta m^2_{\rm sol}}$, where the general discussion presented above
does not apply. In the absence of cancellations, this situation  
corresponds to $(V_L)_{k1}\simeq U_{k1}$, which could arise in certain
models with $V_R$ very close to the identity. In this case, an 
upper bound on the lightest right-handed neutrino mass can
 be derived from taking the trace of 
\begin{equation}
V_R^\dagger D^{-2}_M V_R =\frac{1}{\langle H_u^0\rangle^4}
D^{-1}_Y \hatmass^\dagger  D^{-2 }_Y \hatmass D^{-1}_Y\;,
\end{equation}
which gives
\begin{equation}
\frac{1}{M_1^2}+\frac{1}{M_2^2}+\frac{1}{M_3^2}=
\frac{1}{\langle H_u^0\rangle^4}
\sum_{ij}\frac{|\hatmass_{ij}|^2}{y^2_i y^2_j}\;.
\end{equation}
Therefore, a very conservative bound on $M_1$ is:
\begin{equation}
M_1\leq \frac{y_2^2\langle H_u^0\rangle^2}{ |\hatmass_{22}|}\;,
\label{M1-special}
\end{equation}
where $|\hatmass_{22}|\gsim \sqrt{\Delta m^2_{\rm sol}}$. On the
other hand, an upper bound on $y_2$ can be obtained from
Eq.~(\ref{P-terms}). Keeping the leading term, which in the
absence of cancellations constitutes by itself a lower bound 
on $P_{12}$, we obtain
\begin{eqnarray}
|P_{12}|&\gsim& \left| y_3^2 (V_L)_{13} (V_L)^*_{23} +
y_2^2 (V_L)_{12} (V_L)^*_{22} +y_1^2 (V_L)_{11} (V_L)^*_{21}\right|
\log\left(\frac{\Lambda}{M_3}\right)\nonumber \\
&\simeq& \left| (y_3^2-y^2_2) (V_L)_{13} (V_L)^*_{23} +
(y_1^2-y_2^2) U_{11} U^*_{21}\right|
\log\left(\frac{\Lambda}{M_3}\right)\;,
\label{P12-especial}
\end{eqnarray}
where we have used the unitarity of $V_L$ and the fact that
$(V_L)_{k1}\simeq U_{k1}$. The lowest 
value is reached when $(V_L)_{13} (V_L)^*_{23}\simeq 0$, thus
yielding
\begin{equation}
|P_{12}|\gsim \frac{y_2^2}{3} \log\left(\frac{\Lambda}{M_3}\right)\;.
\end{equation}
Substituting in Eq.~(\ref{M1-special}) finally gives:
\begin{equation}
M_1 \lsim  3|P_{12}| 
      \frac{\langle H_u^0\rangle^2}{\sqrt{\Delta m^2_{\rm sol}}}
\log^{-1} \frac{\Lambda}{M_3}
\;,
\end{equation}
which is comparable in magnitude to the result obtained for generic values
of $V_L$, Eq.~(\ref{M1fromP12}).
Therefore, the bound  for the lightest right-handed neutrino mass 
in terms of ${\rm BR}(\mu \rightarrow e \gamma)$ derived in 
Eq.~(\ref {M1fromBR}) also applies to the special case where 
$|\hatmass_{11}|\ll \sqrt{\Delta m^2_{\rm sol}}$. 

If the neutrino parameters satisfy the relation $(V_L)_{k1}\simeq U_{k1}$,
the effective neutrino mass $\mtilde1$ can be much smaller than 
$\sqrt{\Delta m^2_{\rm sol}}$ without cancellations. Then, being the lepton
asymmetry only weakly washed-out, the observed baryon asymmetry can
be generated even when the absolute lower bound on the lightest right-handed 
neutrino mass, $M_1\gsim 10^9$ GeV, is saturated. As a consequence, following
Eq.~(\ref{BRfromM1}), it would be necessary an improvement in sensitivity
to the process $\mu\rightarrow e\gamma$ of eight orders of magnitude in
order to close the leptogenesis window, which is below the sensitivity of
any planned experiment.

One should note, however, that the lower bound 
Eq.~(\ref{BRfromM1}) is very conservative and can be largely
enhanced by the term proportional to $y_3^2 (V_L)_{13} (V_L)^*_{23}$
in Eq~(\ref{P12-especial}).\footnote{
It is interesting to note that the requirement 
of successful leptogenesis leads to a lower bound on $y_3$, stemming
from the condition $m_3\leq y^2_3 \langle H_u^0 \rangle^2/M_1$.
Therefore, $y_3 \gsim \sqrt{m_3 M_1}/\langle H_u^0 \rangle$,
being $m_3\simeq \sqrt{\Delta m^2_{\rm atm}}$ and $M_1\gsim 10^9$ GeV,
which gives $y_3 \gsim 10^{-3}$. For generic values of $(V_L)_{13}$ and 
$(V_L)_{23}$ the contribution from the largest Yukawa coupling to $P_{12}$ 
can be much larger than the minimal contributions from $y_2$ considered
here, thus yielding much larger rates for ${\rm BR}(\mu\rightarrow e\gamma)$.}
Thus, even though this scenario is the worst 
case scenario for probing leptogenesis, which requires the 
{\it non}-observation of the process $\mu\rightarrow e\gamma$, it is
very favourable for observing a signal in future experiments 
searching for rare decays~\cite{Masiero:2002jn}.

\section{Conclusions}

The see-saw mechanism is perhaps the most elegant explanation
for the small neutrino masses, which in addition provides
a potential solution to the longstanding puzzle of the origin of the 
matter-antimatter asymmetry of the Universe, through the mechanism
of leptogenesis. However, although it is
very appealing theoretically, it suffers the serious disadvantage
of lacking predictability. Furthermore, being the scale of the new
physics presumably very large, it also suffers the disadvantage of
lacking testability. On the other hand, in the supersymmetric
version of the see-saw mechanism, which is probably the most natural
arena to implement it, the high-energy see-saw parameters
leave an imprint on the slepton soft mass matrices through quantum
effects, thus opening a unique opportunity to test the see-saw mechanism 
or the leptogenesis mechanism with low energy observations.

Working under very general and well motivated assumptions, 
namely the absence of cancellations and a hierarchical
pattern in the neutrino Yukawa eigenvalues, we have identified
the scenario yielding the minimal rate for the rare decay
$\mu\rightarrow e\gamma$. In this scenario, the rate depends
essentially on the lightest neutrino Yukawa eigenvalue and on
supersymmetric parameters. Using the experimental constraint
on ${\rm BR}(\mu\rightarrow e\gamma)$ we have derived
an upper bound on the smallest neutrino Yukawa eigenvalue
$y_1\lsim 4\times 10^{-2}$ for typical soft SUSY breaking terms
of 200 GeV and $\tan\beta=10$.

We have shown that this upper bound on the smallest neutrino
Yukawa eigenvalue in turn translates into an upper bound on
the lightest right-handed neutrino mass, $M_1\lsim 5\times 10^{12}$ GeV,
which should be compared with the lower bound required by the thermal
leptogenesis scenario, $M_1\gsim 10^9$ GeV. The upper bound 
derived in this paper scales as ${\rm BR}(\mu\rightarrow e\gamma)^{1/2}$,
therefore,
future improvements in sensitivity to the process $\mu\rightarrow e\gamma$
(and to $\mu-e$ conversion in nuclei) will have important implications
for the thermal leptogenesis scenario if no positive signal is found.
Namely, under the assumption of hierarchical eigenvalues and barring 
cancellations, if supersymmetry is discovered
at the LHC, an improvement in sensitivity 
of six orders of magnitude to ${\rm BR}(\mu\rightarrow e\gamma)$ 
(or seven orders of magnitude to the rate of $\mu-e$ conversion in nuclei)
will suffice to rule out large classes of thermal leptogenesis models based
on the decay of the lightest right-handed neutrino. Possible ways
out are to accept that neutrino parameters take very special values
or to invoke non-minimal scenarios of leptogenesis, such as
leptogenesis induced by the decay of the next-to-lightest 
right-handed neutrino~\cite{Vives:2005ra}.

\section*{Acknowledgements}

AI would like to thank the Yukawa Institute for Theoretical Physics 
for hospitality during the last stages of this work. We are
grateful to Sacha Davidson, Toni Riotto and Tetsuo Shindou 
for useful discussions.
This work was partially supported 
by the DFG cluster of excellence Origin and Structure of the Universe
and by the Graduiertenkolleg ``Particle Physics at the Energy Frontier
of New Phenomena''.

\end{document}